\documentstyle[12pt]{article}
\begin{document}
\title{{\bf STRESS TENSORS \\ FOR INSTANTANEOUS VACUA \\ IN 1+1 DIMENSIONS}
\thanks{Alberta-Thy-02-96, gr-qc/9603005}}
\author{
Don N. Page
\thanks{Internet address:
don@phys.ualberta.ca}
\\
CIAR Cosmology Program, Institute for Theoretical Physics\\
Department of Physics, University of Alberta\\
Edmonton, Alberta, Canada T6G 2J1
}
\date{(1997 June 19)}

\maketitle
\large
\begin{abstract}
\baselineskip 15.6 pt

The regularized expectation value of the stress-energy tensor for a massless
bosonic or fermionic field in 1+1 dimensions is calculated explicitly for the
instantaneous vacuum relative to any Cauchy surface (here a spacelike curve) in
terms of the length $L$ of the curve (if closed), the local extrinsic curvature
$K$ of the curve, its derivative $K'$ with respect to proper distance $x$ along
the curve, and the scalar curvature $R$ of the spacetime:  $T_{00} =  -
\epsilon \pi/(6L^2) - K^2/(24\pi), \: T_{01} = - K'/(12\pi), \: T_{11} = -
\epsilon \pi/(6L^2) - K^2/(24\pi) + R/(24\pi)$, in an orthonormal frame with
the spatial vector parallel to the curve.  Here $\epsilon = 1$ for an untwisted
(i.e., periodic in $x$) one-component massless bosonic field or for a twisted
(i.e., antiperiodic in $x$) two-component massless fermionic field, $\epsilon =
-{1 \over 2}$ for a twisted one-component massless bosonic field, and $\epsilon
= - 2$ for an untwisted two-component massless fermionic field.  The
calculation uses merely the energy-momentum conservation law and the trace
anomaly (for which a very simple derivation is also given herein, as well as
the expression for the Casimir energy of bosonic and fermionic fields twisted
by an arbitrary amount in $R^{D-1} \times S^1$).  The two coordinate and
conformal invariants of a quantum state that are (nonlocally) determined by the
stress-energy tensor are given.  Applications to topologically modified
deSitter spacetimes, to a flat cylinder, and to Minkowski spacetime are
discussed.
\\
\\
\end{abstract}
\normalsize
\baselineskip 15.6 pt

\section*{I.  INTRODUCTION AND MAIN RESULT}

\hspace{.25in}In a globally hyperbolic spacetime of 1+1 dimensions (one
timelike and one spacelike), the trace anomaly
[1--17]
and the conservation law for the regularized expectation value of the
stress-energy-momentum tensor of a massless field determine this stress tensor
throughout the spacetime if it is given on a Cauchy surface (a spacelike
surface, here a one-dimensional curve or line, that each inextendible
nonspacelike curve intersects once and only once; the existence of a Cauchy
surface is equivalent to global hyperbolicity) \cite{DFU,DF,F77,D77,W78}.
However, the stress tensor on a Cauchy line depends on the quantum state.

Here we consider states which are the instantaneous vacua relative to Cauchy
lines.  That is, for any Cauchy line, we consider the state which is the vacuum
with respect to the instantaneous normal-ordered Hamiltonian at the line for
the evolution of the massless field off the line in a gaussian normal
coordinate system in which the time $t$ is the proper time from the Cauchy line
along a timelike geodesic (say of constant comoving spatial coordinate $x$)
that intersects the Cauchy line orthogonally.

In this gaussian normal coordinate system, the spacetime metric in the
neighborhood of the Cauchy line where the coordinate system is nonsingular
(e.g., in the region where the orthogonal timelike geodesics of different $x$
have not crossed each other) has the standard form
 \begin{equation}
 ds^2 = -dt^2 + a^2(t,x)dx^2,
 \label{eq:1}
 \end{equation}
where the Cauchy line is at $t=0$.  I choose the comoving spatial coordinate
$x$ to be proper distance along the Cauchy line (with an unspecified origin),
so on the Cauchy line $t=0$ we have $a=1$.  I shall assume that the topology of
the spacetime is $R^1\times S^1$, where the $R^1$ is the time $t$, so that the
Cauchy line is a closed curve $S^1$ in $x$, and its total proper length (the
period of $x$) will be denoted by $L$, meaning that $(t,x)$ is identified with
$(t,x+L)$.  (For a spacetime with topology $R^2$ in which the Cauchy line has
the topology of $R^1$ and is infinitely long, one may simply set $L=\infty$ in
the resulting formulas.)

A one-dimensional connected manifold (with metric), such as the Cauchy line,
has no intrinsic curvature, so its only intrinsic geometric property is its
total length $L$ and its connectivity (whether it is topologically $S^1$, as
assumed here unless $L = \infty$, or whether it is $R^1$).  However, its
imbedding in the spacetime has a local extrinsic curvature $K(x)$ at each point
$x$, the logarithmic rate of change of the proper distance separation of the
orthogonal geodesics (here the comoving timelike geodesics of constant $x$).
In the gaussian coordinate system above,
 \begin{equation}
 K = {\dot{a} \over a}
 \label{eq:2}
 \end{equation}
for any line of constant $t$, where the overdot denotes a derivative with
respect to the proper time $t$ along the comoving geodesics (which, by the
construction of gaussian normal coordinates, are orthogonal to all $t={\rm
const.}$ lines).  Along the Cauchy line itself, where I have chosen coordinates
so that $a=1$ there, one simply has $K(x)=\dot{a}$.

Any two-dimensional spacetime has its Riemann curvature tensor algebraically
determined purely by the metric tensor $g_{\mu\nu}$ and by the scalar curvature
$R$ as
 \begin{equation}
 R_{\mu\nu\rho\sigma} = {1 \over 2}R
(g_{\mu\rho}g_{\nu\sigma} - g_{\mu\sigma}g_{\nu\rho}),
 \label{eq:3}
 \end{equation}
and in the gaussian normal coordinates above one can readily calculate that the
scalar curvature of the spacetime is
 \begin{equation}
 R = {2\ddot{a} \over a} = 2(\dot{K} + K^2).
 \label{eq:4}
 \end{equation}
(I am using the sign conventions of MTW  \cite{MTW}, except that the sign of
the extrinsic curvature is here chosen to be opposite that of MTW in order that
expanding comoving geodesics correspond to positive extrinsic curvature.)

The instantaneous Hamiltonian on the Cauchy line for a (one-component) real
massless scalar field $\phi$ is simply
 \begin{equation}
 H=\int_{0}^{L}{{1 \over 2}(\pi_{\phi}^2 + \phi'^2) dx},
 \label{eq:5}
 \end{equation}
where $\pi_{\phi}$ is the momentum conjugate to $\phi$ and the prime denotes a
spatial derivative with respect to the proper distance $x$ along the line.
(For the instantaneous Hamiltonian on another $t={\rm const.}$ line, the prime
must mean $a^{-1}d/dx$ in order that it be the derivative with respect to
proper distance there, and $dx$ must be replaced by $a\,dx$ in order that it be
the proper distance interval, so when $a$ is time-dependent the instantaneous
Hamiltonian depends on the time of the line.)  The instantaneous Hamiltonian on
the Cauchy line depends only on the intrinsic properties of the line and not on
its embedding in the spacetime or on the curvature of the spacetime (though of
course the proper distance along the line, and thus also the length of the
line, is induced from the metric of the spacetime).  Thus, on the Cauchy line
itself, the instantaneous vacuum of the normal-ordered Hamiltonian depends only
on the intrinsic properties of that line.

In fact, a short calculation shows that this instantaneous vacuum, for an
untwisted scalar field (one with periodic boundary conditions in the periodic
spatial coordinate $x$), has the configuration-space representation
wavefunctional
 \begin{equation}
 \Psi[\phi(x)] \propto \exp{\left\{{1 \over 4\pi}\int_{0}^{L}dx
 \int_{0}^{L}d\tilde{x} \ln{\left[4\sin^2{{\pi(x-\tilde{x})
 \over L}} \right]} \phi'(x)\phi'(\tilde{x})\right\}},
 \label{eq:6}
 \end{equation}
where the prime on $\phi'(\tilde{x})$ has the obvious meaning of
$d/d\tilde{x}$.  [The zero-mode, $\phi(x)={\rm const.}$, is not damped and is
not normalizable, but this is precisely analogous to the fact that the ground
state of a free nonrelativistic particle in infinite flat space is not
normalizable, because its position is completely indeterminate
\cite{SS,FPhD,DF}.  With the zero-mode being in its nonnormalizable
zero-momentum ground state, it makes no contribution to the stress tensor,
though it does make the expectation value of the two-point function
$\phi(x)\phi(\tilde{x})$ divergent.  Only quantities invariant under a constant
shift of $\phi$, such as the stress tensor, can be well defined and finite.]
However, we shall not need the explicit form of the wavefunctional here, which
is only given to show directly that the instantaneous vacuum state on a Cauchy
line does not depend on the extrinsic curvature of the line or the curvature of
the spacetime.

Now that the notation and situation have been established, the main result, to
be established below, can be simply stated:  On the Cauchy line, and in the
frame in which $d/dt$ is the unit timelike vector and $d/dx$ is the orthonormal
spacelike vector, the regularized stress-energy tensor expectation value
$T_{\mu\nu} \equiv \langle \hat{T}_{\mu\nu} \rangle$ (not bothering with
angular brackets around this expectation value for simplicity, since I will
rarely need a symbol for the regularized stress-energy tensor operator
$\hat{T}_{\mu\nu}$ itself) for $N_b$ massless bosonic field components (e.g.,
$N_b$ real scalar fields, or ${1 \over 2}N_b$ complex scalar fields; $N_b$
counts the number of one-particle states for a given momentum) and for $N_f$
massless fermionic field components (e.g., $N_f$ real Majorana-Weyl spinor
fields or ${1 \over 2}N_f$ two-component complex fermion-antifermion spinor
fields) has the orthonormal covariant components of energy density
 \begin{equation}
 T_{tt} = - {N_b + {1 \over 2} N_f \over 24\pi}
 \left( K^2 + \epsilon{4 \pi^2 \over L^2} \right),
 \label{eq:7}
 \end{equation}
energy flux or momentum density
 \begin{equation}
 T_{tx} = T_{xt} = - {N_b + {1 \over 2} N_f \over 24\pi} 2 K',
 \label{eq:8}
 \end{equation}
and pressure
 \begin{equation}
 T_{xx} = - {N_b + {1 \over 2} N_f \over 24\pi}
 \left( K^2 - R + \epsilon{4 \pi^2 \over L^2} \right),
 \label{eq:9}
 \end{equation}
where $\epsilon = 1$ for untwisted (i.e., periodic in $x$) massless bosonic
fields or for a standard twisted (i.e., antiperiodic in $x$ \cite{I,I2})
massless fermionic fields, $\epsilon = -{1 \over 2}$ for a twisted massless
bosonic field, $\epsilon = - 2$ for an untwisted massless fermionic field,
$\epsilon = 1 - 6\chi + 6\chi^2$ for complex bosonic fields $\phi$ with the
boundary condition $\phi(x+L) = e^{2 \pi i \chi}\phi(x)$ with $0 \leq \chi \leq
1$, and $\epsilon = - 1 + 6\chi - 6\chi^2$ for complex fermionic fields $\psi$
with the boundary condition $\psi(x+L) = e^{2 \pi i \chi}\psi(x)$, again with
$0 \leq \chi \leq 1$.  (The ordinary untwisted fields correspond to $\chi = 0$
or $\chi = 1$, and the standard twisted fields correspond to $\chi = {1 \over
2}$.  In the theory of closed strings \cite{GSW}, the Ramond boundary condition
\cite{R} gives untwisted fermionic fields, and the Neveu-Schwarz boundary
condition \cite{NS} gives standard twisted fermionic fields on the string
worldsheet.  Twisting with $2\chi$ nonintegral occurs in orbifolds and in
four-dimensional fermionic string constructions \cite{DHVW,KLT,LT}, or to
coupling a charged field to a flat $U(1)$ connection that is topologically
nontrivial over the $S^1$ corresponding to the periodic $x$.)  If there are
$N_b$ bosonic field components with various twistings $\chi_i$ and $N_f$ field
components with various twistings $\chi_j$, the general formula for $\epsilon$
is
 \begin{equation}
 \epsilon = (N_b + {1 \over 2} N_f)^{-1}
 \left[ \sum_{i=1}^{N_b}{(1 - 6\chi_i + 6\chi_i^2)}
 - \sum_{j=1}^{N_f}{(1 - 6\chi_j + 6\chi_j^2)} \right].
 \label{eq:9b}
 \end{equation}
Since the $\chi$'s are restricted to lie between 0 and 1 inclusive, one can
immediately see that $\epsilon$ is restricted to lie between $-2$ (its value
for purely untwisted fermionic fields) and 1 (its value for purely untwisted
bosonic fields and/or standard twisted fermionic fields).

This stress tensor expectation value is nonlocally determined, because the
quantum state (the instantaneous vacuum) is nonlocally defined, but in 1+1
dimensions the nonlocal dependence is on the only nonlocal intrinsic geometric
property of the Cauchy line, namely its length $L$ (along with the type of
boundary condition for the field).  The dependence on the local quantities of
the extrinsic curvature of the line and the scalar curvature of the spacetime
arise from the definition of the regularized stress tensor operator.  In higher
dimensions, the regularized stress tensor operator would remain local, but the
instantaneous vacuum state would in general have a much more complicated
nonlocal dependence on the intrinsic geometry of the spatial Cauchy
hypersurface.

\section*{II.  DERIVATION OF THE TRACE ANOMALY}

\hspace{.25in}In order to derive the regularized stress tensor expectation
values given above, we use the energy-momentum conservation law and the trace
anomaly.  For the benefit of those who, like me, may have difficulty in
remembering the precise formula for the trace anomaly even in the simple case
of 1+1 dimensions, let me give a simple derivation of it from certain
easily-remembered facts that I will not derive.  (Others may wish to skip this
digression, though it gives a simple case of the method to be used to calculate
the stress tensor on a general Cauchy line.  After coming up with it, I found
that somewhat similar simple derivations had been given in \cite{CF,D77,W78},
and doubtless every other element of this derivation is also somewhere in the
literature, though I have not seen all the parts put together previously in
such a simple way that the value of the trace anomaly can literally be worked
out in one's head in the middle of the night.)

The trace of the regularized stress-energy tensor of a massless field in 1+1
dimensions is a local covariant analytic scalar function of the metric and its
derivatives that has no dimensional parameters in it and which vanishes in flat
spacetime.  Its dimensionality of inverse length squared then forces it to be
$\alpha R$ for some pure number $\alpha$ \cite{DDI}, which might in principle
depend on the number and types of fields, but which does not depend on the
global boundary conditions of the fields.  (The dimensionality requires the
function to have two derivatives in it, and multiples of the scalar curvature
are the only local covariant scalar functions of the metric and its derivatives
with precisely two derivatives.  If one used a covariant expression involving
more derivatives of the metric, such as $R^{;\mu}_{;\mu}$, one would have to
take a root to get the dimension right, but this would not be analytic in the
limit of flat spacetime.)

Consider a spacetime which is flat $S^1\times R^1$ for $t<0$ and the unit
deSitter metric for $t>0$, with metric
 \begin{equation}
 ds^2 = -dt^2 + [\theta(-t) + \theta(t) \cosh^2{t}]dx^2
 \label{eq:10}
 \end{equation}
(where $\theta$ is the step function that is 0 for negative argument, ${1 \over
2}$ for zero argument, and 1 for positive argument) and with $x$ identified
with period $L = 2\pi$.  The instantaneous vacuum for the Cauchy line $t=0$ is
obviously the same as for any line of constant $t<0$ and gives the static
vacuum for the flat region of the spacetime.  (It also is the same as the
instantaneous vacuum for any line with constant $t>0$, but I will not need this
fact here.)  This vacuum has, in the flat region $t<0$, a constant Casimir
energy density \cite{C,BD,F} $\rho_C$ and an equal pressure $P_C = \rho_C$
(since the trace is zero in the flat region).  There is no energy flux in this
situation, which is symmetrical under $x \rightarrow -x$.

When one crosses the Cauchy line to $t$ slightly positive, the trace anomaly
$\alpha R$ can change the pressure.  However, the energy-momentum conservation
law prevents the energy density from changing discontinuously across the Cauchy
line.  (It would jump, as we shall find in the more general case below, if the
pressure had a term that were a delta function in time, but this could arise
only if the curvature, and hence the trace anomaly, had a delta function in it.
 This would occur if the intrinsic curvature were not the same on the two sides
of the Cauchy line, but it is zero on both sides in the metric above, since the
Cauchy line is a geodesic for both metrics.)  Thus the energy density is
$\rho_C$ also on the deSitter side of the Cauchy line.

The static vacuum for the flat region of the spacetime ($t \leq 0$), evaluated
on the geodesic Cauchy line at $t=0$, is given by a path integral over field
configurations, on the $\tau\leq 0$ half of the Euclidean (i.e.,
positive-definite-metric) cylinder obtained by setting $t = -i\tau$ and
analytically continuing $\tau$ to real negative values, which match the
argument of the wavefunctional on the $\tau = 0$ Cauchy line and which vanish
asymptotically as $\tau\rightarrow -\infty$.  Similarly, the deSitter-invariant
vacuum for $t \geq 0$, evaluated on the geodesic Cauchy line at $t = 0$, is
given by a path integral over the hemisphere, bounded by this line, that is the
Euclidean analytic continuation of the deSitter Lorentzian spacetime metric one
gets by setting $t = i({\pi \over 2} - \theta)$ and taking $\theta$ to be a
real spherical polar angle in the range between $0$ (where the hemisphere
closes off at its `north pole') and ${\pi \over 2}$ (at the equator where the
match is made to the Lorentzian deSitter metric at its geodesic Cauchy line).
(This is only true if $x$ has period $L = 2\pi$, which is the period one gets
by the standard construction of the unit deSitter metric as the metric induced
on a unit timelike hyperboloid in flat 2+1 dimensional spacetime.  Regularity
of fields at the north pole requires a bosonic field to be untwisted and a
fermionic field to be twisted in order to give a deSitter-invariant vacuum by
this construction, since increasing $x$ by its period of $2\pi$ corresponds to
a rotation of $2\pi$ radians about the pole.  One might have expected this even
from the Lorentzian 1+1 dimensional deSitter spacetime, viewed as a timelike
hyperboloid embedded in flat 2+1 dimensional spacetime, since going once around
the hyperboloid corresponds to a rotation of $2\pi$ in the enveloping flat
spacetime, and a rotation of $2\pi$ reverses the sign of fermionic, but not
bosonic, fields.)

The metric on this hemispheric analytic continuation of the deSitter spacetime
is conformally related to the metric on the flat Euclidean half-cylinder.
Furthermore, field configurations on the hemisphere that are regular at the
north pole correspond to field configurations on the cylinder that vanish
exponentially as $\tau\rightarrow -\infty$.  Therefore, since the action of a
massless field in a two-dimensional spacetime is conformally invariant, the
path integrals over the hemisphere and half-cylinder give the same values or
amplitudes [given by Eq. (\ref{eq:6}) above for an untwisted scalar field] for
the same field configurations on the bounding Cauchy lines.  Thus the
instantaneous vacuum of an untwisted bosonic or a twisted fermionic field on
the $t=0$ Cauchy line, which is the static vacuum for $t<0$, is also the
deSitter invariant vacuum for $t>0$.

One can readily check this by a one-page calculation, which I shall not bother
repeating here (and which I could not do in my head in the middle of the
night), showing that if one takes any Cauchy line of uniform extrinsic
curvature, $K'=0$ (including, of course, the cases $K=0$ in which the Cauchy
lines are geodesic, but not restricting oneself to the $K'=0$ lines of constant
$t$), in the complete 1+1 deSitter spacetime in which $a = \cosh{t}$ for all
negative as well as positive time, then the resulting instantaneous vacuum is
the same, as it should be in order to have the boost and conformal invariance
of the deSitter spacetime.  (There is no mixing of positive and negative
frequency modes between any two Cauchy lines of uniform extrinsic curvature in
1+1 dimensional deSitter spacetime.)

Therefore, the instantaneous vacuum (for a periodic bosonic field or an
antiperiodic fermionic field) in the $t>0$ region of the metric (\ref{eq:10})
has the same invariance as the deSitter spacetime, and its regularized
stress-energy tensor must be simply proportional to the metric.  This gives a
trace of twice the negative of the energy density, namely $-2\rho_C$.  Since
the scalar curvature of the unit-scale deSitter region of the metric above is
$R=2$, the coefficient of $R$ in the trace anomaly is $\alpha = -\rho_C$, minus
the Casimir energy density in the periodic space of length $2\pi$ for $t<0$.

Next, one remembers that the Casimir energy density in a space periodic in one
dimension can readily be obtained from a corresponding thermal energy density
and pressure \cite{T}:  The Casimir effect can be obtained by a path integral
over the Euclidean space of topology $R^1\times S^1$ obtained by making the
Lorentzian time imaginary (i.e., $\tau = it$ real), as discussed above.  If the
coordinates $\tau$ and $x$ are interchanged, this Euclidean space is precisely
the same as the $S^1\times R^1$ space that one obtains by assigning an
imaginary period of magnitude $2\pi$ to time in flat Lorentzian spacetime and
then taking this time to be imaginary.  This is the space on which the path
integral gives the thermal energy density $\rho_T$ of flat spacetime at a
temperature $T = {1 \over 2\pi}$, the inverse of the period of the imaginary
time
[31--37].  The antiperiodicity in $x$ that we needed above for the fermionic
field becomes precisely the antiperiodicity in imaginary time needed for a
thermal state of fermionic fields.  Because of making the opposite coordinates
of the Euclidean $R^1\times S^1$ imaginary in going to the unbounded Lorentzian
thermal case in contrast to going to the periodic Lorentzian (Casimir) vacuum
case, and because the two Euclidean orthonormal components $T_{\tau\tau}$ and
$T_{xx}$ have equal magnitudes but opposite signs in order to give a trace of
zero, one gets $\rho_C = -\rho_T$ and hence $\alpha = \rho_T$.  Finally, an
elementary calculation gives, for a one-component bosonic field (e.g., a real
scalar field) at temperature $T = {1 \over 2\pi} $,
 \begin{equation}
 \alpha_s = \rho_{T_s} =
 \int_{-\infty}^{\infty}{{dk \over 2\pi}{|k|
 \over e^{|k|/T}-1}} = {\pi T^2 \over 6} = {1 \over 24\pi}.
 \label{eq:11}
 \end{equation}

One gets the same result \cite{DU} for a two-component massless fermionic field
(e.g., neutrinos plus anti-neutrinos), with the integral for each component
being half as large:
 \begin{equation}
 \alpha_f = \rho_{T_f} = 2\int_{-\infty}^{\infty}
 {{dk \over 2\pi}{|k|
 \over e^{|k|/T}+1}} = {\pi T^2 \over 6} = {1 \over 24\pi}.
 \label{eq:11b}
 \end{equation}

Since the trace anomaly is a local quantity, it does not depend on the
periodicity properties of the fields.  Thus it is ${R \over 24\pi}$ for either
a one-component bosonic field (untwisted or twisted) or a two-component
fermionic field (twisted or untwisted).  Of course, for a two-component bosonic
field, such as a complex scalar field, it is twice as large.  Therefore, for
$N_b$ bosonic components and $N_f$ fermionic components, the trace anomaly is
 \begin{equation}
 T^{\mu}_{\mu} = \alpha R
 = {1 \over 24\pi} (N_b + {1 \over 2} N_f) R.
 \label{eq:11c}
 \end{equation}

\section*{III.  CASIMIR ENERGY DENSITY OF TWISTED \\ FIELDS IN ARBITRARY
DIMENSIONS}

\hspace{.25in}For use below, it is also of interest to calculate the Casimir
energy density for twisted bosonic and untwisted fermionic fields.  In a
periodic one-dimensional space of length $L$, it has been given by Isham
\cite{I} as ${\pi \over 12L^2}$ for a standard twisted scalar field [$\phi(x+L)
= - \phi(x)$], which is $-{1 \over 2}$ the value one gets by the argument above
for an untwisted scalar field (after scaling the length from $2\pi$ to $L$).
Citing results from \cite{DHI,I,DB} in 1+3 dimensions, Avis and Isham \cite{AI}
note that ``twice the (regularized) self-energy of a scalar field plus the
spinor's self-energy sums to zero:  a typical supersymmetry result.''  This
indeed gives ${\pi \over 12L^2}$ for a twisted massless real scalar field, and
it also gives $-2$ times the value for an untwisted scalar field, or ${\pi
\over 3L^2}$, for an untwisted massless two-component fermionic field in 1+1
dimensions, which is what I shall take here, though Davies and Unruh \cite{DU}
seem to state a value half as large, and Birrell and Davies \cite{BD} seem to
state a value twice as large (possibly from assuming a different number of
components for the fermionic field).

Another possibility one can consider is the case in which one has a complex
field which, when one goes once around the period of $x$, returns to its
original value multiplied by the phase $e^{2 \pi i \chi}$, with $0 \leq \chi
\leq 1$.  (Alternatively, the complex field may be left periodic but coupled to
a flat but nontrivial $U(1)$ gauge field whose integral around the period of
$x$, when multiplied by the charge coupling constant of the field, gives a
phase of $2 \pi i \chi$, up to an integral multiple of $2 \pi$ that has no
effect.)  For a complex scalar field with this partial twisting in 1+1
dimensions, Dowker and Banach \cite{DB} find that the Casimir energy density is
$2(1 - 6\chi + 6\chi^2)$ times the untwisted ($\chi = 0$) density of $- {\pi
\over 6L^2}$ a single real scalar field.  (The factor of 2 is for the two real
components of the complex scalar field.)  If the supersymmetry result stated by
Isham \cite{I} is valid when extended to this case, that the fermionic Casimir
energy density is the opposite of that of the same number of bosonic components
with the same periodicity conditions, then a two-component fermionic field
which gets multiplied by the phase $e^{2 \pi i \chi}$ when one goes around the
loop given by the coordinate $x$ of proper length $L$ would have a Casimir
energy density of $2\pi (1 - 6\chi + 6\chi^2)/(6L^2)$.  This does give the same
values as above in the untwisted case ($\chi =0$) and in the standard twisted
case ($\chi = {1 \over 2}$), so in this paper I shall assume that it is
correct.  Then the Casimir energy density in 1+1 dimensions for $N_b$ bosonic
field components with various twistings $\chi_i$ and $N_f$ field components
with various twistings $\chi_j$ is
 \begin{equation}
 \rho_C = - {4\pi^2\alpha\epsilon \over L^2}
 = - {\pi \over 6 L^2}
 \left[ \sum_{i=1}^{N_b}{(1 - 6\chi_i + 6\chi_i^2)}
 - \sum_{j=1}^{N_f}{(1 - 6\chi_j + 6\chi_j^2)} \right].
 \label{eq:11d}
 \end{equation}

Incidentally, if one combines the supersymmetry result \cite{AI} with the trick
of Toms \cite{T} for getting the Casimir energy density in a periodic space of
$S^1$ length $L$ (with periodic boundary conditions for bosonic fields and with
antiperiodic boundary conditions for fermionic fields) as the negative of the
thermal pressure in an infinite space with temperature $1/L$, then one can
readily give it for both untwisted and standard twisted bosonic and fermionic
fields in a flat spacetime of dimension $D$ and topology $S^1 \times R^{D-1}$:
 \begin{equation}
 \rho_C = \pi^{-D/2}\Gamma(D/2)\zeta(D)L^{-D}
 [ - N_{ub} + N_{uf} + (1 - 2^{1-D})(N_{tb} - N_{tf})],
 \label{eq:A}
 \end{equation}
where $\Gamma(D/2)$ is the ordinary gamma function,
 \begin{equation}
 \zeta(D) \equiv \sum_{n=1}^{\infty}{{1 \over n^D}}
 \label{eq:B}
 \end{equation}
is the ordinary Riemann zeta function, $L$ is the (constant) length of the
periodic $S^1$ spatial dimension, $N_{ub}$ is the number of untwisted (periodic
in the $S^1$) bosonic field components (degrees of freedom or number of
one-particle states for a given momentum), $N_{uf}$ is the number of untwisted
fermionic field components, $N_{tb}$ is the number of twisted (antiperiodic in
the $S^1$) bosonic field components, and $N_{tf}$ is the number of twisted
fermionic field components.

Furthermore, one can readily generalize the results of Dowker and Banach
\cite{DB} (say by using the method of images for the Green function \cite{DHI})
to get the Casimir energy density of $N_b$ bosonic and $N_f$ fermionic field
components that are paired into complex components that are twisted by the
arbitrary phase angles $2\pi\chi_i$:
 \begin{equation}
 \rho_C = - \pi^{-D/2}\Gamma(D/2)L^{-D}
 \sum_{i=1}^{N_b + N_f}
 {(-1)^{F_i}\sum_{n=1}^{\infty}
 {{\cos{2\pi n \chi_i} \over n^D}}},
 \label{eq:C}
 \end{equation}
where $F_i=0$ if the $i$th component is bosonic and $F_i=1$ if the $i$th
component is fermionic.  This essentially just replaces the Riemann zeta
function in Eq. (\ref{eq:A}) with the final sum in Eq. (\ref{eq:C}), which
obviously is $\zeta(D)$ for $\chi_i = 0$ or $\chi_i = 1$ and is $- (1 -
2^{1-D})\zeta(D)$ for $\chi_i = {1 \over 2}$.  When the total number of
dimensions of the spacetime is even, $D = 2m$ for some integer $m$, one may
evaluate this sum explicitly \cite{AS,GR} to get
 \begin{eqnarray}
 \rho_C &=& {(-2\pi)^m \over 2m(2m-1)!! L^D}
 \sum_{i=1}^{N_b + N_f}{(-1)^{F_i}B_{2m}(\chi_i)} \\
 &=& {(-2\pi)^m \over 2m(2m-1)!! L^D}
 \sum_{i=1}^{N_b + N_f}{(-1)^{F_i}
 \sum_{k=0}^{2m}{\left (\matrix{2m\cr k\cr}\right )
 B_k \chi_i^{2m-k}}}
 \label{eq:D}
 \end{eqnarray}
in terms of Bernoulli polynomials $B_{2m}(\chi_i)$ or of Bernoulli numbers
$B_k$, for $0 \leq \chi_i \leq 1$.  This of course reduces to Eq.
(\ref{eq:11d}) for $m=1$ ($D=2$).

Of course, given the uniform Casimir energy density $\rho_C$ in a periodic flat
spacetime, the entire stress-energy may be obtained immediately by Toms'
thermal correspondence \cite{T} as diagonal in the appropriate orthonormal
frame, with the pressure in the periodic direction being $(D-1)\rho_C$ and the
pressure in each of the $D-2$ transverse directions being $- \rho_C$.

\section*{IV.  DERIVATION OF THE STRESS TENSOR \\
OF AN INSTANTANEOUS VACUUM}

\hspace{.25in}Having completed this digression of calculating the trace anomaly
of $N_b$ bosonic and $N_f$ fermionic massless field components in 1+1
dimensions and the Casimir energy density in a spacetime of any dimension that
is static and periodic in one of the spatial dimensions, we may return to
calculating the stress tensor expectation value of the instantaneous vacuum on
an arbitrary Cauchy line in an arbitrary 1+1 dimensional spacetime.  For this
we perform a similar trick of replacing the part of the spacetime below the
line ($t<0$) with a flat spacetime such that the extrinsic curvature is zero on
that side.  That is, in the metric (\ref{eq:1}) we keep the correct form for
$a(t,x)$ for $t>0$ but set $a(t,x) = 1$ for $t<0$.  This keeps the intrinsic
geometry, $ds^2 = dx^2$, the same on both sides of the Cauchy line, but it
gives a discontinuity in the extrinsic curvature $K$ and hence a delta-function
contribution in the scalar curvature $R$ of the spacetime.  Thus we can no
longer argue that the energy density is continuous across the Cauchy line, but
we can still readily use the energy-momentum conservation law, along with the
trace anomaly, to calculate what the stress tensor is just above the Cauchy
line.

This law, $T^{\mu\nu}_{\;\;\;\; ;\nu} = 0$, is simplest to solve in null
coordinates in which the metric has the form
 \begin{equation}
 ds^2 = - e^{2\sigma} du dv,
 \label{eq:12}
 \end{equation}
with $\sigma$ being a function of the two null coordinates $u$ and $v$.  In
terms of $\sigma$, the only nonzero Christoffel symbols in the $(u,v)$
coordinate basis are
 \begin{equation}
 \Gamma^u_{uu} = 2 \sigma_{,u}\; , \;\;
 \Gamma^v_{vv} = 2 \sigma_{,v}\; ,
 \label{eq:13}
 \end{equation}
and the scalar curvature is
 \begin{equation}
 R = - 2 \sigma^{;\mu}_{;\mu} = 8 e^{-2\sigma} \sigma_{,uv}\; ,
 \label{eq:14}
 \end{equation}
where the comma denotes a partial derivative, so the trace anomaly is
 \begin{equation}
 T^\mu_\mu = - 4 e^{-2\sigma} T_{uv} = \alpha R
 = {1 \over 24\pi} (N_b + {1 \over 2}N_f) R
 = {1 \over 3\pi} (N_b + {1 \over 2}N_f) e^{-2\sigma} \sigma_{,uv}\; .
 \label{eq:15}
 \end{equation}

Now an explicit calculation of $T^{\mu\nu}_{\;\;\;\; ;\nu} = 0$ in these null
coordinates, making use of the trace anomaly, readily shows that
\cite{DFU,DF,F77,D77,W78}
 \begin{equation}
 T_{uu} = 2 \alpha
 \left[ \sigma_{,uu} - \sigma_{,u}\sigma_{,u} + U(u) \right],
 \label{eq:16}
 \end{equation}
 \begin{equation}
 T_{uv} = T_{uv} = - 2 \alpha \sigma_{,uv}\; ,
 \label{eq:17}
 \end{equation}
 \begin{equation}
 T_{vv} = 2 \alpha
 \left[ \sigma_{,vv} - \sigma_{,v}\sigma_{,v} + V(v) \right],
 \label{eq:18}
 \end{equation}
where $U$ is an arbitrary function of the one null coordinate $u$, and $V$ is
an arbitrary function of the other null coordinate $v$.

In the flat spacetime for $t<0$, we can choose $u = t - x$ and $v = t + x$, so
$\sigma = 0$ there.  The instantaneous vacuum on the Cauchy line is the static
vacuum for $t<0$, which has $T_{uv} = T_{vu} =0$ and $\rho_C = T_{uu} + T_{vv}
= 4 \alpha U(u) = 4 \alpha V(v) = - 4 \pi^2 \alpha \epsilon / L^2$, with
$\epsilon$ being defined by Eq. (\ref{eq:9b}) for a general set of massless
fields with various periodicity conditions, or by the preceding discussion in
special cases.  Then Eqs. (\ref{eq:16})--(\ref{eq:18}) gives the stress tensor
in the region $t>0$ in terms of the derivatives of $\sigma$ there.

To calculate the result explicitly, one needs to transform the metric
(\ref{eq:1}) from the $(t,x)$ coordinate system to the null coordinate system
$(u,v)$ in which the metric has the form (\ref{eq:12}).  To the order needed,
one finds that
 \begin{equation}
 u \approx t - {1 \over 2} K t^2 - x,
 \label{eq:19}
 \end{equation}
 \begin{equation}
 v \approx t - {1 \over 2} K t^2 + x,
 \label{eq:20}
 \end{equation}
 \begin{equation}
 \sigma \approx \ln{a} \approx K t +
 ( - {1 \over 2} K^2 + {1 \over 4} R) t^2.
 \label{eq:21}
 \end{equation}
Here $K$ means $K(x)$, the value of $K(t,x) = \dot{a}/a$ at $t=0$ [i.e., $K$ as
a function purely of $x \approx (v-u)/2$ at $u+v = 0$], rather than the value
of $K(t,x)$ at the point of interest.

Now, remembering the implicit $x$ (but not $t$) dependence of $K$ in Eqs.
(\ref{eq:19})--(\ref{eq:21}) when one differentiates $\sigma$, one gets the
following expressions [with $K'$ denoting $dK(x)/dx$] on the Cauchy line $t=0$
(or $u+v=0$):
 \begin{equation}
 \sigma_{,uu} = {1 \over 8} R - {1 \over 2} K',
 \label{eq:22}
 \end{equation}
 \begin{equation}
 \sigma_{,uv} = \sigma_{,uv} = {1 \over 8} R,
 \label{eq:23}
 \end{equation}
 \begin{equation}
 \sigma_{,vv} = {1 \over 8} R + {1 \over 2} K'.
 \label{eq:24}
 \end{equation}
Inserting these back into Eqs. (\ref{eq:16})--(\ref{eq:18}), along with $U(u) =
V(v) = - \pi^2 \epsilon / L^2$, gives the following components of the
stress-energy tensor on the Cauchy line in the null coordinate system:
 \begin{equation}
 T_{uu} = \alpha
 \left({1 \over 4} R - {1 \over 2} K^2 - K'
 - \epsilon{2 \pi^2\over L^2} \right),
 \label{eq:25}
 \end{equation}
 \begin{equation}
 T_{uv} = T_{vu} = - {1 \over 4} \alpha R,
 \label{eq:26}
 \end{equation}
 \begin{equation}
 T_{vv} = \alpha
 \left({1 \over 4} R - {1 \over 2} K^2 + K'
 - \epsilon{2 \pi^2\over L^2} \right).
 \label{eq:27}
 \end{equation}
When these components are transformed back to the $(t,x)$ coordinates (which
are orthonormal coordinates on the Cauchy line), and when one uses Eq.
(\ref{eq:11c}) for the coefficient $\alpha$ of $R$ in the trace anomaly, one
gets the stress-energy tensor components listed above in Eqs.
(\ref{eq:7})--(\ref{eq:9}).  This completes the derivation of the main result.

\section*{V.  COVARIANT FORMS FOR THE STRESS \\ TENSOR AND CONFORMAL
INVARIANTS}

\hspace{.25in}It is of interest to write the stress-energy tensor in three
explicitly covariant forms, in terms of the latter two of which one may find
two nonlocal conformal invariants of the quantum state.  For example, the
general expressions (\ref{eq:16})--(\ref{eq:18}) for the components of a
conserved stress-energy tensor with trace $\alpha R$ in 1+1 dimensions can be
written in terms of an auxiliary scalar field $\Phi$ in the covariant form
\cite{W}
 \begin{equation}
 T_{\mu\nu} = \Phi_{;\mu}\Phi_{;\nu}
 - {1 \over 2}g_{\mu\nu}(\nabla \Phi)^2
 + 2 \kappa (\Phi_{;\mu\nu} - g_{\mu\nu}
{\,\lower0.9pt\vbox{\hrule \hbox{\vrule height 0.2 cm \hskip 0.2 cm \vrule
height 0.2 cm}\hrule}\,}\Phi),
 \label{eq:C1}
 \end{equation}
where $\kappa \equiv \sqrt{\alpha/2}$, $\: (\nabla \Phi)^2 \equiv \Phi^{;\mu}
\Phi_{;\mu}$, and
 \begin{equation}
 {\,\lower0.9pt\vbox{\hrule \hbox{\vrule height 0.2 cm \hskip 0.2 cm \vrule
height 0.2 cm}\hrule}\,}\Phi
 \equiv \Phi^{;\mu}_{;\mu} = - \kappa R.
 \label{eq:C2}
 \end{equation}
Here I have chosen the normalization of $\Phi$ so that if one sets $\kappa = 0$
(or, more realistically, considers a field $\Phi$ with large derivatives so
that the quadratic terms in $\Phi$ in Eq. (\ref{eq:C1}) dominate over the
linear terms in $\Phi$), then the stress-energy tensor (\ref{eq:C1}) has the
standard form for a classical scalar field $\phi = \Phi$.

When the metric is written in the form of Eq. (\ref{eq:12}), one may write
 \begin{equation}
 \Phi = 2 \kappa (\sigma + f),
 \label{eq:C3}
 \end{equation}
where ${\,\lower0.9pt\vbox{\hrule \hbox{\vrule height 0.2 cm \hskip 0.2 cm
\vrule height 0.2 cm}\hrule}\,}f \equiv
 - 4 e^{-2\sigma}f_{,uv} = 0$, so $f$ must have the form of a function of the
null coordinate $u$ plus another function of the null coordinate $v$, $f(u,v) =
p(u) + q(v)$.  Then one can easily calculate that in Eqs. (\ref{eq:16}) and
(\ref{eq:18}),
 \begin{equation}
 U(u) = f_{,uu} + f_{,u}f_{,u} = {d^2 p\over du^2}
 + \left( {d{p} \over du} \right)^2
 = e^{-p}{d^{2} \over d{u^2}}{e^p},
 \label{eq:C4}
 \end{equation}
 \begin{equation}
 V(v) = f_{,vv} + f_{,v}f_{,v} = {d^2 q\over dv^2}
 + \left( {d{q} \over dv} \right)^2
 = e^{-q}{d^{2} \over d{v^2}}{e^q}.
 \label{eq:C5}
 \end{equation}

Although a given $\Phi$ determines a unique stress-energy tensor by Eq.
(\ref{eq:C1}), the reverse is not true, in that $\Phi$ is not uniquely
determined by $T_{\mu\nu}$, at least locally.  That is, for fixed functions
$U(u)$ and $V(v)$ (uniquely determining and uniquely determined by $T_{\mu\nu}$
once the null coordinates $u$ and $v$, and hence the conformal factor
$e^{2\sigma}$ in the metric (\ref{eq:12}), are fixed for a given geometry),
there are two-parameter local solutions of Eqs. (\ref{eq:C4}) and (\ref{eq:C5})
for both $p(u)$ and $q(v)$.  [One of the resulting four parameters cancels when
one adds $p(u)$ to $q(v)$ to get $f$, and one of the remaining three parameters
corresponds to the obvious invariance of the stress-energy tensor (\ref{eq:C1})
under adding a constant to $\Phi$.]  In particular, for the vacuum (Casimir)
values $U(u) = V(v) = - \pi^2 \epsilon / L^2$, one has the general local
solution for $f$ being
 \begin{equation}
 f = f_0 + \ln{\sin{{\pi\sqrt{\epsilon} (u - u_0) \over L}}}
 + \ln{\sin{{\pi\sqrt{\epsilon} (v - v_0) \over L}}},
 \label{eq:C6}
 \end{equation}
with the three arbitrary constants $f_0$, $u_0$, and $v_0$.  Note that for
$\epsilon < 1$, $f$ cannot have the periodicity $(u,v) \equiv (u-L,v+L)$ of the
spacetime unless $u_0$ and $v_0$ are taken to $\pm i\infty$, in which case one
gets the one-parameter sets of global solutions $f = \pm i\pi
\sqrt{\epsilon}(u+v+c)/L$ for the single constant $c$.  For $\epsilon = 1$, the
general local three-parameter solution (\ref{eq:C6}) has the same periodicity
as the coordinates and so is a global solution, but even there one cannot have
$u_0$ or $v_0$ real if the solution is to be finite everywhere, and so, just as
for other positive values of $\epsilon$, one needs $f$ and hence $\Phi$ to be
complex, a consequence of the negativity of the Casimir energy density in that
case.

A second covariant way to get a conserved stress-energy tensor with trace
$\alpha R$ in 1+1 dimensions is to start with a nowhere null or vanishing
closed divergenceless one-form ${\mbox{\boldmath $ \omega$}} = \omega_{\mu}{\bf
d}x^{\mu}$, i.e., one obeying ${\mbox{\boldmath $ \omega\!\cdot\!\omega $}}
\equiv \omega_{\mu}\omega^{\mu} \neq 0$, ${\mbox{\boldmath $d\omega $}} =
{\mbox{\boldmath $0$}}$ (or $\omega_{[\mu ; \nu]} = 0$), and ${\mbox{\boldmath
$ \delta \omega $}} \equiv \ast {\mbox{\boldmath $ d \!$}} \! \ast \!
{\mbox{\boldmath $ \! \omega $}} = 0$ (or $\omega_{\mu}^{\;\; ;\mu} = 0$), so
that locally, but not necessarily globally, ${\mbox{\boldmath $\omega $}} =
{\mbox{\boldmath $d$}}h$ for some scalar field $h$ obeying
${\,\lower0.9pt\vbox{\hrule \hbox{\vrule height 0.2 cm \hskip 0.2 cm \vrule
height 0.2 cm}\hrule}\,}h = 0$.  In terms of null coordinates $(u,v)$, any
closed divergenceless one-form can be written in terms of two functions, $P$
and $Q$, each of one of the two null coordinates:
 \begin{equation}
 {\mbox{\boldmath $ \omega $}}
 = P(u){\bf d}u + Q(v){\bf d}v.
 \label{eq:C7}
 \end{equation}
In this representation, the condition that the one-form be nowhere null or
vanishing is the condition that $P(u)$ and $Q(v)$ are both nowhere zero.  Then
construct the scalar field
 \begin{equation}
 F = \ln{(c \,{\mbox{\boldmath $ \omega\!\cdot\!\omega $}})}
 \equiv \ln{(c \,\omega_{\mu}\omega^{\mu})}
 = - 2\sigma + \ln{P(u)} + \ln{Q(v)} - \ln{(-4c)}
 \label{eq:C8}
 \end{equation}
for an arbitrary nonzero constant $c$ [e.g., $c = {\rm sgn}({\mbox{\boldmath $
\omega\!\cdot\!\omega $}})$].  This scalar field automatically obeys
${\,\lower0.9pt\vbox{\hrule \hbox{\vrule height 0.2 cm \hskip 0.2 cm \vrule
height 0.2 cm}\hrule}\,}F = R$.  Now one can write the stress-energy tensor as
 \begin{equation}
 T_{\mu\nu} = \alpha
 \{2\omega_{\mu}\omega_{\nu} - F_{;\mu\nu}
 + {1 \over 2}F_{;\mu}F_{;\nu}
 + [-{\mbox{\boldmath $ \omega\!\cdot\!\omega $}}
 + {\,\lower0.9pt\vbox{\hrule \hbox{\vrule height 0.2 cm \hskip 0.2 cm \vrule
height 0.2 cm}\hrule}\,}F
 - {1 \over 4}(\nabla F)^2] g_{\mu\nu}\}.
 \label{eq:C9}
 \end{equation}
A comparison with Eqs. (\ref{eq:16}) and (\ref{eq:18}) shows that now
 \begin{equation}
 U(u) = P^2 + {3 \over 4P^2}\left( {d{P} \over du} \right)^2
 - {1 \over 2P}{d^{2}P \over du^2}
 = y^{-4} + y^{-1}{d^{2}{y} \over du^2},
 \label{eq:C10}
 \end{equation}
 \begin{equation}
 V(v) = Q^2 + {3 \over 4Q^2}\left( {d{Q} \over dv} \right)^2
 - {1 \over 2Q}{d^{2}Q \over dv^2}
 = z^{-4} + z^{-1}{d^{2}{z} \over dv^2},
 \label{eq:C11}
 \end{equation}
where $y(u) = P^{-1/2}$ and $z(v) = Q^{-1/2}$, so
 \begin{equation}
 {\mbox{\boldmath $ \omega $}} = {{\bf d}u \over y^2(u)}
 + {{\bf d}v \over z^2(v)}.
 \label{eq:C12}
 \end{equation}
Since $P(u)$ and $Q(v)$ are nowhere zero, $y$ and $z$ are everywhere finite,
though they are imaginary if $P$ and $Q$ are negative, or complex if $P$ and
$Q$ themselves are pure imaginary, which would also give a real stress-energy
tensor (\ref{eq:C9}).

Analogously to the situation with the stress-energy tensor (\ref{eq:C1}), so
the stress-energy tensor (\ref{eq:C9}) is uniquely determined by the closed
divergenceless one-form {\boldmath $ \omega $} but does not uniquely determine
it, at least locally.  For example, $P(u)$ and $Q(v)$ uniquely determine $U(u)$
and $V(v)$, but there is a two-parameter set of local solutions of Eqs.
(\ref{eq:C10}) and (\ref{eq:C11}) for each of them for given $U(u)$ and $V(v)$.
 In particular, for the vacuum (Casimir) values $U(u) = V(v) = - \pi^2 \epsilon
/ L^2$, one has the general local solutions for $P$ and $Q$ being
 \begin{equation}
 P(u) = {\pm i\pi \sqrt{\epsilon}/L
 \over \cosh{\psi_0} + \sinh{\psi_0}
 \sin{[2\pi\sqrt{\epsilon}(u - u_0)/L]}}.
 \label{eq:C13}
 \end{equation}
 \begin{equation}
 Q(v) = {\pm i\pi \sqrt{\epsilon}/L
 \over \cosh{\tilde{\psi}_0} + \sinh{\tilde{\psi}_0}
 \sin{[2\pi\sqrt{\epsilon}(v - v_0)/L]}},
 \label{eq:C14}
 \end{equation}
where $u_0$, $\psi_0$, $v_0$, and $\tilde{\psi}_0$ are four arbitrary constants
(possibly complex), and where each of the two arbitrary signs may be chosen
independently.  However, if the spacetime has the periodicity $(u,v) \equiv
(u-L,v+L)$ that we have been assuming, so that $U(u+L) = U(u)$ and $V(v+L) =
V(v)$, then unless $\epsilon = 1$, we must have $\psi_0 = 0$ and
$\tilde{\psi}_0 = 0$, so that one has only the four discrete (zero-parameter)
solutions
 \begin{equation}
 P(u) = \pm Q(v) = \pm i\pi \sqrt{\epsilon}/L,
 \label{eq:C15}
 \end{equation}
again with the two signs independent.

For a spacetime of topology $R^1\times S^1$, the one-form {\boldmath $ \omega
$} that determines the stress-energy tensor (\ref{eq:C9}) gives two conformal
invariants of the quantum state, each a nonlocal functional of the
geometry and of the stress-energy tensor field:
 \begin{equation}
 {\cal E}_u = \epsilon +
 \left( {1 \over \pi}
 \int_{u_0}^{u_0 + L}{\omega_u du} \right)^2,
 \label{eq:C16}
 \end{equation}
 \begin{equation}
 {\cal E}_v = \epsilon +
 \left( {1 \over \pi}
 \int_{v_0}^{v_0 + L}{\omega_v dv} \right)^2,
 \label{eq:C17}
 \end{equation}
where the integral for ${\cal E}_u$ is taken along the null line $v = {\rm
const.}$, and the integral for ${\cal E}_v$ is taken along the null line $u =
{\rm const}$.  (Of course I could have simply written each integrand above as
the one-form {\boldmath $ \omega $}, but the coordinate forms above shows more
explicitly which null coordinate is to varied for the corresponding integral.)

It is important to note that in the cases in which the two-parameter set of
local solutions for either $P$ or $Q$ are also global solutions in the topology
$R^1\times S^1$ (which occurs only when $\epsilon = 1$ and only when either the
left-moving modes or the right-moving modes, respectively, are in their ground
state in some coordinate and conformal frame), then any choice out of the
possible solutions of Eq. (\ref{eq:C10}) or (\ref{eq:C11}) for $P$ or $Q$ gives
the same value for the corresponding conformal invariant, namely ${\cal E}_u =
0$ or ${\cal E}_v = 0$.  That is, in these special cases in which there are
many global choices for ${\mbox{\boldmath $ \omega $}} = P(u){\bf d}u +
Q(v){\bf d}v$ that lead to the same stress-energy tensor (\ref{eq:C9}), each of
these choices gives the same values in Eqs. (\ref{eq:C16}) and (\ref{eq:C17})
for the invariants ${\cal E}_u$ and ${\cal E}_v$.  This ambiguity in {\boldmath
$ \omega $} occurs for the vacuum state of untwisted bosonic fields and
standard twisted fermionic fields in deSitter spacetime, which is the
instantaneous vacuum for {\it any} geodesic Cauchy line through any point:  one
gets different choices for {\boldmath $ \omega $} by setting $P(u)$ and $Q(v)$
to have the constant value given by Eq. (\ref{eq:C15}) in each of various
different null coordinate systems $(u,v)$ such that $-u = v = x$ along the
corresponding geodesic Cauchy line with proper distance $x$, and these choices,
when transformed back to a single coordinate system, give $P(u)$ and $Q(v)$ of
the forms given by Eqs. (\ref{eq:C13}) and (\ref{eq:C14}).

${\cal E}_u$ and ${\cal E}_v$ are invariants not only under a change of the
null coordinates $u$ and $v$ to new null coordinates $\tilde{u}(u)$ and
$\tilde{v}(v)$ and hence a corresponding shift in the metric (\ref{eq:12}) of
$\sigma$ to $\tilde{\sigma} = \sigma - {1 \over 2}\ln{{d{\tilde{u}} \over du}}
- {1 \over 2}\ln{{d{\tilde{v}} \over dv}}$, but also under an arbitrary
conformal transformation (Weyl rescaling that may vary arbitrarily with
position) that changes $\sigma(u,v)$ to an arbitrary new function of $u$ and
$v$.  I have chosen the constant terms in Eqs. (\ref{eq:C16}) and
(\ref{eq:C17}) so that ${\cal E}_u$ and ${\cal E}_v$ are both zero if and only
if the quantum state is the instantaneous vacuum of a set of quantum fields
[with $\epsilon$ determined by Eq. (\ref{eq:9b}) from the boundary conditions
of how twisted the various fields are] for at least one Cauchy line (which can
be put into the form $\tilde{u} + \tilde{v} = 0$ for some choice of null
coordinates $\tilde{u}$ and $\tilde{v}$) and for at least one choice of the
conformal factor in the metric (in particular, for the choices that gives
$\tilde{\sigma} = {\rm const.}$).  For a state which is not the instantaneous
vacuum for any choice of Cauchy line and for any conformally transformed
metric, ${\cal E}_u$ and/or ${\cal E}_v$ are positive.  [Neither one can be
negative for the stress-energy tensor arising from a quantum state of a set of
fields with the corresponding value of $\epsilon$, though one can certainly
choose an (imaginary) {\boldmath $ \omega $}, and hence an artificial conserved
stress-energy tensor (\ref{eq:C9}) with the correct trace but with ${\cal E}_u$
and/or ${\cal E}_v$ arbitrarily negative.  Analogously, one can write down in
Minkowski spacetime a covariantly constant (and hence conserved) artificial
stress-energy tensor that has a constant negative energy density everywhere,
but one cannot get such a tensor as the expectation value of the regularized
stress-energy tensor operator in any quantum state.]  I have arbitrarily chosen
the normalization of ${\cal E}_u$ and ${\cal E}_v$ so that for any combination
of untwisted massless bosonic fields and the standard twisted massless
fermionic fields (giving $\epsilon = 1$ and a negative Casimir energy density
for the ground state on a spacetime geometry that has a static flat metric on
$R^1\times S^1$), the excited states with zero stress-energy tensor on this
geometry give ${\cal E}_u = 1$ and ${\cal E}_v = 1$.

Of course, there are an infinite number of invariants (under coordinate and
conformal or Weyl transformations of the metric) of a quantum state of massless
fields in 1+1 dimensions, but the only independent ones that are determined
purely by the stress-energy tensor are ${\cal E}_u$ and ${\cal E}_v$ given
above.  One can readily see this, because if one chooses null coordinates
$\tilde{u}\propto\int{\omega_u du}$ and $\tilde{v}\propto\int{\omega_v dv}$ but
normalized so that the periodicity is
$(\tilde{u},\tilde{v})\equiv(\tilde{u}-L,\tilde{v}+L)$, and then makes a
conformal or Weyl transformation so that $\tilde{\sigma} = 0$ in this
coordinate system, then the stress-energy tensor has only the two constant
components $T_{\tilde{u}\tilde{u}}$ and $T_{\tilde{v}\tilde{v}}$, and these two
constants determine only two invariants.  In terms of these two constants, the
invariants given above are
 \begin{equation}
 {\cal E}_u = \epsilon +
 {L^2 \over 2\alpha\pi^2} T_{\tilde{u}\tilde{u}},
 \label{eq:C18}
 \end{equation}
 \begin{equation}
 {\cal E}_v = \epsilon +
 {L^2 \over 2\alpha\pi^2} T_{\tilde{v}\tilde{v}}.
 \label{eq:C19}
 \end{equation}
In particular, if one has a static state for a single untwisted bosonic field
(one with $\epsilon = 1$ and $\alpha = {1 \over 24\pi}$) in which
$T_{\tilde{u}\tilde{u}} = T_{\tilde{v}\tilde{v}} = {E \over 2L}$, and one
further sets the length of the $S^1$ to be $L = 2\pi$, then ${\cal E}_u = {\cal
E}_v = 1 + 12 E$, with the ground state having the energy $E = - {1 \over 12}$.

Other independent coordinate and conformal invariants of the quantum state do
not affect the stress-energy tensor.  For example, in the static flat geometry
on $R^1\times S^1$, the quantum state which is the vacuum excited by precisely
two particles in the left-moving mode $e^{-2\pi i u/L}$ with wavelength $L$ has
the same stress-energy tensor as the quantum state which is the vacuum excited
by precisely one particle in the left-moving mode $e^{-4\pi i u/L}$ with
wavelength $L/2$, though these states are surely not equivalent to each other
under any coordinate and/or conformal transformations.

A third and very similar covariant way to get a conserved stress-energy tensor
with trace $\alpha R$ in 1+1 dimensions is to start with a nowhere null or
vanishing conserved traceless symmetric second-rank tensor {\boldmath $ \tau $}
(i.e., a tensor with covariant components $\tau_{\mu\nu}$ obeying
$Tr\,{\mbox{\boldmath $\tau $}}^2\equiv\tau^{\alpha\beta}\tau_{\alpha\beta}\neq
0$, $\tau^{\mu\nu}_{\;\;\;\; ;\nu} = 0$, $Tr\,{\mbox{\boldmath $\tau
$}}\equiv\tau^{\alpha}_{\;\; \alpha} = 0$, and $\tau_{\mu\nu} =
\tau_{\nu\mu}$), which in terms of null coordinates $(u,v)$ must have the form
 \begin{equation}
 {\mbox{\boldmath $ \tau $}}
 = \tau_{uu}(u){\bf d}u\otimes{\bf d}u
 + \tau_{vv}(v){\bf d}v\otimes{\bf d}v
 \label{eq:C20}
 \end{equation}
where $\tau_{uu}(u)$ and $\tau_{vv}(v)$ are each a function only of the
corresponding null coordinate and are each everywhere nonzero.  Then construct
the scalar function
 \begin{equation}
 {\cal F}
 = {1 \over 4}\ln{[Tr({\mbox{\boldmath $\tau $}}^2)^2]}
 \equiv {1 \over 4}
 \ln{[(\tau^{\alpha\beta}\tau_{\alpha\beta})^2]}
 \label{eq:C21}
 \end{equation}
and the stress-energy tensor
 \begin{equation}
 T_{\mu\nu} = \tau_{\mu\nu} +
 \alpha \{{1 \over 2}{\cal F}_{;\mu}{\cal F}_{;\nu}
 - {\cal F}_{;\mu\nu}
 + [{\,\lower0.9pt\vbox{\hrule \hbox{\vrule height 0.2 cm \hskip 0.2 cm \vrule
height 0.2 cm}\hrule}\,}{\cal F}
 - {1 \over 4}(\nabla {\cal F})^2] g_{\mu\nu}\}.
 \label{eq:C22}
 \end{equation}

If one compares this to Eq. (\ref{eq:C9}), one sees that
 \begin{equation}
 {\mbox{\boldmath $ \tau $}}
 = \alpha(2{\mbox{\boldmath $ \omega \!\otimes\! \omega
 - \omega\!\cdot\!\omega \, g $}}),
 \label{eq:C23}
 \end{equation}
or, in component form,
 \begin{equation}
 \tau_{\mu\nu} = \alpha(2\omega_{\mu}\omega_{\nu}
 - \omega^{\alpha}\omega_{\alpha}g_{\mu\nu}),
 \label{eq:C24}
 \end{equation}
and that ${\cal F}$ is the same as $F$, up to an additive constant that drops
out of the derivatives appearing in Eqs. (\ref{eq:C9}) and (\ref{eq:C22}).
Furthermore, one can readily calculate that, analogous to Eqs. (\ref{eq:C10})
and (\ref{eq:C11}), $\tau_{uu}(u)$ and $\tau_{vv}(v)$ obey the nonlinear
second-order differential equations
 \begin{equation}
 4\alpha\tau_{uu}\tau_{uu;uu} - 5\alpha\tau_{uu;u}\tau_{uu;u}
 - 8\alpha\tau_{uu}\tau_{uu}\tau_{uu}
 + 8T_{uu}\tau_{uu}\tau_{uu} = 0,
 \label{eq:C25}
 \end{equation}
 \begin{equation}
 4\alpha\tau_{vv}\tau_{vv;vv} - 5\alpha\tau_{vv;v}\tau_{vv;v}
 - 8\alpha\tau_{vv}\tau_{vv}\tau_{vv}
 + 8T_{vv}\tau_{vv}\tau_{vv} = 0,
 \label{eq:C26}
 \end{equation}

One might consider $\tau_{\mu\nu}$ to be the ``classical'' (traceless)
stress-energy tensor corresponding to the full quantum stress-energy tensor
$T_{\mu\nu}$ (by which is meant here, as throughout this paper, the expectation
value $\langle \hat{T}_{\mu\nu} \rangle$ of the regularized stress-energy
tensor operator $\hat{T}_{\mu\nu}$ in the quantum state of the fields), though
the second-order character of Eqs. (\ref{eq:C25}) and (\ref{eq:C26}) mean that
$\tau_{\mu\nu}$ is not locally uniquely determined by $T_{\mu\nu}$.  However,
in a spacetime with the topology $R^1\times S^1$, as we have been assuming in
this paper, $\tau_{\mu\nu}$ is globally uniquely determined by $T_{\mu\nu}$ if
$\epsilon\neq 1$ or if both the invariants ${\cal E}_u$ and ${\cal E}_v$ are
nonzero.

We can also readily write these coordinate and conformal invariants in terms of
the tensor $\tau_{\mu\nu}$ in the following way:  Construct the auxiliary flat
metric
 \begin{equation}
 d\tilde{s}^2 = \tau_{\mu\nu}dx^{\mu}dx^{\nu}
 = \tau_{uu}(u)du^2 + \tau_{vv}(v)dv^2,
 \label{eq:C27}
 \end{equation}
which can have any signature if $\epsilon > 0$ (but which is necessarily
positive definite if $\epsilon < 0$), and using it evaluate the squared
spacetime interval $\tilde{s}^2_u$ (the square of the geodesic distance if the
points are spacelike separated, or minus the square of the proper time
separation if the points are timelike separated) between the points $(u,v)$ and
$(u+L,v)$, and the squared spacetime interval $\tilde{s}^2_v$ between the
points $(u,v)$ and $(u,v+L)$.  Then a comparison with Eqs. (\ref{eq:C16}) and
(\ref{eq:C17}) shows that
 \begin{equation}
 {\cal E}_u = \epsilon +
 {\tilde{s}^2_u \over 2\alpha\pi^2},
 \label{eq:C28}
 \end{equation}
 \begin{equation}
 {\cal E}_v = \epsilon +
 {\tilde{s}^2_v \over 2\alpha\pi^2}.
 \label{eq:C29}
 \end{equation}
These formulas show that it might have been more natural to define these
invariants to be a factor of $2\alpha\pi^2$ larger than what I have defined,
which would have also made them scale linearly with the stress-energy tensor if
one increased the number of fields (and hence $\alpha$), but I have chosen a
normalization so that if $\epsilon = 1$, states with zero energy flux and
nonpositive Casimir energy density in a spatially-periodic static flat
spacetime have the invariants between 0 (the value for the ground state) and 1
(the value when the energy density is zero).

\section*{VI.  EXAMPLES OF INSTANTANEOUS VACUUM \\ STRESS TENSORS IN SIMPLE
SPACETIMES}

\hspace{.25in}Now we can apply the formulas (\ref{eq:7})--(\ref{eq:9}) or
(\ref{eq:25})--(\ref{eq:27}) for the stress-energy tensor of an instantaneous
vacuum (which necessarily has ${\cal E}_u = {\cal E}_v = 0$) to a few simple
examples.  We start with the 1+1 dimensional deSitter spacetime and one type of
modification of it.  For example, if we apply the formulas above to the unit
deSitter metric of Eq. (\ref{eq:10}) (with $R = 2$) for a Cauchy line of
constant $t>0$, we find that $L = 2\pi \cosh{t}$, $K = \tanh{t}$, $K' = 0$, so
for the untwisted one-component bosonic or twisted two-component fermionic
field one gets the deSitter-invariant stress tensor $T_{\mu\nu} = {1 \over
48\pi}R g_{\mu\nu}$ as the stress tensor of the instantaneous vacuum
corresponding to any Cauchy line of constant $K$.  There is a unique state
giving this stress tensor, the deSitter invariant state of the massless field
\cite{BD78} (provided one allows the homogeneous mode to be in its
nonnormalizable state of zero conjugate momentum \cite{SS,FPhD,DF}).

However, if we apply the stress tensor calculation to a modified deSitter
metric with the same local metric components as Eq. (\ref{eq:10}), but with $x$
having a period $2\pi \lambda$ different from $2\pi$ for a constant $\lambda$
different from one, then the length of the $t=0$ Cauchy line is $L = 2\pi
\lambda$, and the stress tensor of the instantaneous untwisted bosonic or
twisted fermionic field vacuum of this line is not deSitter invariant (at least
for for $\epsilon = 1$).  For example, the null radiation components are
 \begin{equation}
 T_{uu} = T_{vv}
 = {1 \over 2}\alpha (1 - \epsilon\lambda^{-2}),
 \label{eq:28}
 \end{equation}
and these are not invariant under boosts, unless $\epsilon = \lambda^2$.  (For
$N_b$ bosonic field components and $N_f$ fermionic field components all
partially twisted by the angle $2\pi\chi$, evaluating $\epsilon$ by Eq.
(\ref{eq:9b}) gives the null radiation components as
 \begin{equation}
 T_{uu} = T_{vv} = {N_b(\lambda^2 - 1 + 6 \chi - 6 \chi^2)
 + N_f({1 \over 2}\lambda^2 + 1 - 6 \chi + 6 \chi^2)
 \over 48\pi\lambda^2},
 \label{eq:28b}
 \end{equation}
which can be made zero for any positive $\lambda < 1$ by two particular choices
of $\chi$, symmetrically arranged about $\chi = {1 \over 2}$, but for $\lambda
> 1$ there is no real choice of $\chi$ which will make the null radiation
components zero for $N_b$ and/or $N_f$ positive.)

If one writes the unit local deSitter metric given by the $t>0$ part of Eq.
(\ref{eq:16}) in null coordinates $u = {\rm gd}\:t - x \equiv
\sin^{-1}{\tanh{t}} - x$ and $v = {\rm gd}\:{t} + x \equiv \sin^{-1}{\tanh{t}}
+ x$, where gd is the Gudermannian function, the metric becomes
 \begin{equation}
 ds^2 = {-du dv \over \cos^2{[(u+v)/2]}}.
 \label{eq:29}
 \end{equation}
The modification of the global deSitter structure to give $x$ period $2\pi
\lambda$, rather than $2\pi$, means that $(u, v)$ is identified with
$(u-2\pi\lambda, v+2\pi\lambda)$.  Then Eqs. (\ref{eq:16})--(\ref{eq:18}) imply
that the null radiation components given by Eq. (\ref{eq:28}) or (\ref{eq:28b})
are the same in this coordinate system over the entire modified deSitter
spacetime.  These are also the radiation components of the instantaneous vacuum
with respect to any $t={\rm const.}$ Cauchy line, since all these vacua are the
same.  [However, one must note that the null coordinates defined by Eqs.
(\ref{eq:19}) and (\ref{eq:20}) near a $t={\rm const.} \neq 0$ Cauchy line are
scaled to be a factor $\cosh{t}$ larger than the global null coordinates in Eq.
(\ref{eq:29}).]  The constancy of these covariant null components of the stress
tensor does not imply the constancy of the orthonormal components, and indeed
the trace of the square of the traceless or radiation part of the stress-energy
tensor,
 \begin{equation}
 \tilde{T}_{\mu\nu} \equiv
 (T_{\mu\nu}-{1 \over 2}T^\rho_\rho g_{\mu\nu}),
 \label{eq:30}
 \end{equation}
depends on the time $t={\rm gd}^{-1}{[(u+v)/2]}\equiv
\tanh^{-1}{\sin{[(u+v)/2]}}$:
 \begin{equation}
 \tilde{T}^{\mu\nu} \tilde{T}_{\mu\nu}
 = 2 \alpha^2 (1 - \epsilon\lambda^{-2})^2 \cosh^{-4}{t}
 = 2 \alpha^2 (1 - \epsilon\lambda^{-2})^2 \cos^4{[(u+v)/2]}.
 \label{eq:31}
 \end{equation}
Thus the instantaneous vacuum of any of the $t={\rm const.}$ Cauchy surfaces
has a stress tensor which is not only frame dependent (not invariant under
boosts) but is also inhomogeneous (in time), unless $\epsilon = \lambda^2$.
(Remember that  $\epsilon \leq 1$.)

For a generic $\lambda$ this is not at all surprising, since the period (in
$x$) of the oscillations of $t$ of the spatial geodesics $t =
\tanh^{-1}{[\gamma\sin{(x-x_0)}]}$ for arbitrary $\gamma$ is $2\pi$, so if $x$
does not have a period that is a integer multiple of this, then the geodesics
(except for the one with $\gamma = 0$) do not close in one period of $x$, and
the global structure of the spacetime does not have the $SO(2,1)$ invariance of
the standard deSitter spacetime.  Nevertheless, for $\lambda$ equal to an
integer $n$ (larger than one, in order not to be the standard deSitter
spacetime), one simply has an $n$-fold covering spacetime of the standard
deSitter spacetime, so the geodesics do close in one period of $x$ and one has
a spacetime with the same $SO(2,1)$ invariance as the standard deSitter
spacetime.  Yet the stress-energy tensor of any of the instantaneous vacua
corresponding to these geodesics does not share this $SO(2,1)$ invariance.  [In
fact the vacua corresponding to geodesics of different $\gamma$ in a fixed
coordinate system, while being related by $SO(2,1)$ transformations, are not
identical but have stress tensors with the null components $T_{uu}$ and
$T_{vv}$ at a common intersection point $(t=0, x=x_0)$ scaled down and up
respectively by the squared boost factor ${1+\gamma \over 1-\gamma}$.]

One may define an `instantaneous radiation energy' $\tilde{E}$ on a Cauchy line
to be the integral over the line of the local energy density $\tilde{T}_{tt}$
of the traceless or radiation part of the stress-energy tensor expectation
value in an arbitrary quantum state:
 \begin{equation}
 \tilde{E} = \int_{0}^{L}{dx \tilde{T}_{tt}}
 = \int_{0}^{L}{dx {1 \over 2}(T_{tt} + T_{xx})}
 = \left\langle :\!H\!: \right\rangle
 - {4 \pi^2 \alpha \epsilon \over L}
 + \alpha \int_{0}^{L}{dx ({1 \over 2}R - K^2)},
 \label{eq:32}
 \end{equation}
the expectation value $\left\langle :\!H\!: \right\rangle$ of the
normal-ordered instantaneous Hamiltonian $H$ of Eq. (\ref{eq:5}), shifted by a
function of the intrinsic and extrinsic properties of the line and the
enveloping spacetime.  [Here $\tilde{T}_{tt}$ is the time-time component in the
orthonormal frame given by the line, i.e., in a coordinate system in which the
metric near the line has the form given by Eq. (\ref{eq:1}) with $a=1$ on the
line.]

Since $\left\langle :\!H\!: \right\rangle$ is bounded below by its value of
zero for the instantaneous vacuum for the corresponding Cauchy line, so is the
instantaneous radiation energy $\tilde{E}$ bounded below by its value
$\tilde{E_0}$ for the instantaneous vacuum:
 \begin{equation}
 \tilde{E} = \left\langle :\!H\!: \right\rangle + \tilde{E_0}
 \geq \tilde{E_0} =
 - {4 \pi^2 \alpha \epsilon \over L}
 + \alpha \int_{0}^{L}{dx ({1 \over 2}R - K^2)}.
 \label{eq:33}
 \end{equation}
For a geodesic Cauchy line in the unit modified deSitter spacetime with $x$
having period $2\pi n$ for some integer $n>1 \geq \epsilon$, one gets
 \begin{equation}
 \tilde{E_0}
 = 2 \pi \alpha \left( {n^2 - \epsilon \over 12 n} \right) > 0.
 \label{eq:34}
 \end{equation}

Thus all states in one of the $n$-fold coverings of the standard deSitter
spacetime have positive instantaneous radiation energy $\tilde{E}$.
But since the components of the traceless or radiation part of the
stress-energy tensor are not invariant under boosts (at least if they are
nonzero and finite), there is no quantum state of a massless field in this
covering of deSitter spacetime with a finite deSitter-invariant stress-energy
tensor expectation value.  (One might think that one could get one by averaging
the instantaneous vacua for the geodesic Cauchy lines of different $\gamma$
over the group of boosts that transform one into another, but this group is
noncompact, and the resulting state, if well defined at all, would have an
infinite stress tensor.)

After writing this section, I found that it is simply a rediscovery of what
Davies and Fulling had pointed out twenty years ago \cite{DF}, with only the
rather trivial extension here to arbitrarily twisted fields and the addition of
the demonstration that there is {\em no} deSitter-invariant state with finite
stress tensor in the $n$-fold covering of deSitter spacetime.  However, a
reminder of these facts may be useful to a forgetful person such as myself.

A second simple example to which we can apply our formulas is (for
concreteness) an untwisted one-component bosonic field or a twisted
two-component fermionic field in the flat ($R=0$) metric on $S^1\times R^1$,
 \begin{equation}
 ds^2 = -dt^2 + d\varphi^2,
 \label{eq:35}
 \end{equation}
in which $\varphi$ is given a period $2\pi$ [i.e., the point given by the
coordinates $(t,\varphi)$ is identified with the point given by the coordinates
$(t,\varphi+2\pi)$].  Here if we take the Cauchy line to be one of the closed
geodesic lines $t={\rm const.}$, the instantaneous vacuum is the standard
static vacuum with energy density and pressure given by the Casimir value
$\rho_C =  - {1 \over 24\pi}$, as derived above.

But suppose we instead take the Cauchy line to be a smooth extrinsically curved
spacelike line $t=t(\varphi)$ with
 \begin{equation}
 \tilde{t} \equiv {d{t} \over d{\varphi}}
 \label{eq:36}
 \end{equation}
having a magnitude less than unity everywhere (but a nonzero magnitude over at
least a nonzero interval of $\varphi$, so that the line is indeed not
geodesic), and with $t(2\pi) = t(0)$ and $\tilde{t}(2\pi) = \tilde{t}(0)$.  The
proper length along the line is then $dx = \sqrt{1-\tilde{t}^2} d\varphi$, so
the total length of the Cauchy line is
 \begin{equation}
 L = \int_{0}^{2\pi}{d\varphi \sqrt{1-\tilde{t}^2}} \; < 2\pi.
 \label{eq:37}
 \end{equation}
The extrinsic curvature of the line is
 \begin{equation}
 K = (1-\tilde{t}^2)^{-3/2}\tilde{\tilde{t}} \equiv
 \left[ 1 -  \left( {d{t} \over d\varphi} \right)^2 \right]^{-3/2}
 {d^2 t \over d\varphi^2}.
 \label{eq:38}
 \end{equation}

Since the length of the Cauchy line is less than that of the closed geodesics
$t={\rm const.}$, and since the extrinsic curvature contributes negatively to
the local energy density of the instantaneous vacuum of the line, the energy
density is everywhere lower than the Casimir value $\rho_C =  - {1 \over
24\pi}$.  Thus the corresponding instantaneous radiation energy (or
instantaneous total energy, since here where the trace is zero the traceless or
radiation part of the stress tensor is the entire stress tensor),
 \begin{equation}
 \tilde{E_0} = - {\pi \over 6 L} - {1 \over 24\pi}
 \int_{0}^{2\pi}{K^2 \sqrt{1-\tilde{t}^2} d\varphi},
 \label{eq:39}
 \end{equation}
is {\it lower} than the Casimir energy, $-{1 \over 12}$, of a geodesic Cauchy
line.

Because the static vacuum, the instantaneous vacuum for a geodesic Cauchy line
in this case, is supposed to be the ground state of this system, it might seem
surprising that a different state, namely the instantaneous vacuum of a curved
Cauchy line, could have a lower instantaneous radiation or total energy.
However, the resolution of this apparent paradox is that the integral for the
instantaneous energy is adding up different components of the stress-energy
tensor (namely, those orthogonal to the curved Cauchy line) than those that go
into the Casimir energy corresponding to the static Killing vector
${\mbox{\boldmath $ \xi $}} = {d \over dt}$.  The energy corresponding to
{\boldmath $ \xi $} for the instantaneous vacuum of a Cauchy line is
 \begin{eqnarray}
 E &=& -\int{\xi^\mu T_\mu^\nu \epsilon_{\nu\varrho}
 dx^\varrho}
 = \int_{0}^{L}{\xi^\mu T_{\mu 0} dx} \nonumber \\
 &=& \int_{0}^{2\pi}{[(1-\tilde{t}^2)^{-1/2} T_{00}-\tilde{t}
(1-\tilde{t}^2)^{-1/2} T_{10}]\sqrt{1-\tilde{t}^2} d\varphi} \nonumber \\
 &=& - {\pi^2 \over 3 L^2}
 - {1 \over 24\pi}\int_{0}^{2\pi}{(K^2 + 2 \tilde{t} K') d\varphi}
 = - {\pi^2 \over 3 L^2}
 + {1 \over 24\pi}\int_{0}^{2\pi}{K^2 d\varphi} \nonumber \\
 &=& \!\!{1\over 12}\left\{\!-\!\left(\int_{0}^{2\pi}
 {{d\varphi \over 2\pi}
 \sqrt{1 \!\! - \!\! \left( {d{t} \over d\varphi} \right)^2}} \, \right)^{-2} +
 \int_{0}^{2\pi}{{d\varphi \over 2\pi}\left[ 1 \!\! - \!\!  \left( {d{t} \over
 d\varphi} \right)^2 \right]^{-3}
 \left( {d^2 t \over d\varphi^2}\right)^2 }\right\},
 \label{eq:40}
 \end{eqnarray}
using 0 and 1 for the timelike and spacelike orthonormal components,
respectively, in the frame given by the Cauchy line, and performing an
integration by parts after remembering that the prime on $K'$ is a derivative
with respect to the proper length $x$ rather than with respect to the angular
coordinate $\varphi$ that has period $2\pi$.

Although it is not obvious to me directly from this expression that the Killing
energy $E$ for a general instantaneous vacuum is bounded below by the Casimir
value $-{1 \over 12}$ of the static vacuum, as it must be by general arguments,
one can see that to lowest order in $\tilde{t}$ it is.  A Fourier expansion of
$t(\varphi)$ shows that to quadratic order all terms in $E + {1 \over 12}$ are
positive, except the $m = \pm 1$ terms, which vanish to quadratic order but
give a positive fourth-order contribution.

Nevertheless, the fact that one can make the instantaneous radiation energy
density arbitrarily negative over the whole Cauchy line (by making the length
of the line arbitrarily short; the $K^2$ term also contributes negatively but
must be zero somewhere on a smooth closed line in this flat spacetime, since
the $x$-integral of $K$ itself is zero by the trivial holonomy of the closed
curves) translates into the fact that one can make even the $\varphi$-density
of the Killing energy arbitrarily negative over all but an arbitrarily small
range of the $\varphi$ integration.  That is, even though the total Killing
energy must be no less than $-{1 \over 12}$ for the flat $S^1\times R^1$
spacetime in which the period of the $S^1$ is $2\pi$ (or no less than $-{1
\over 12\lambda}$ if the period of the $S^1$ were $2\pi \lambda$), the excess
energy can be concentrated into an arbitrarily small region, leaving almost all
of the space with lower energy density than the ground state.

As a third example, consider the flat Minkowski spacetime on the topology
$R^2$, with the metric
 \begin{equation}
 ds^2 = -dt^2 + dz^2,
 \label{eq:41}
 \end{equation}
where both $t$ and $z$ range between $-\infty$ and $+\infty$.  Again, for
concreteness and simplicity, take the standard untwisted one-component bosonic
field or twisted two-component fermionic field.  In this case a Cauchy line
$t(z)$ across the spacetime would have infinite length $L$, unless it consisted
of an infinite number of segments of opposite sign of
 \begin{equation}
 \tilde{t} \equiv {d{t} \over d{z}}
 \label{eq:42}
 \end{equation}
that are sufficiently nearly null [$\tilde{t}$ sufficiently near $\pm 1$, such
as $\tilde{t} = \tanh{(z^2 \sin{z})}$] to make the total length finite, but
then the integral of $K^2$ in the total Killing energy would diverge.  When the
total length $L$ is indeed infinite, the contribution to the Killing energy for
the section of the Cauchy line with $z_1 \leq z \leq z_2$ is
 \begin{equation}
 E(z_1, z_2) = {1 \over 24\pi} \int_{z_1}^{z_2}{dz \left[ \left(
 {d{p} \over dz} \right)^2 - {d^2 \over dz^2}{p^2} \right]},
 \label{eq:43}
 \end{equation}
where
 \begin{equation}
 p \equiv {\tilde{t} \over \sqrt{(1 - \tilde{t}^2)}}
 \equiv \left[ 1 - \left({d{t} \over dz} \right)^2 \right]^{-1/2}
 {d{t} \over d{z}}.
 \label{eq:44}
 \end{equation}
The integrand can be negative for an arbitrarily long range of $z$, for example
by having $K \equiv {d{p} \over dz}$ constant, but if this range in which the
integrand is negative is extended to $z = \pm \infty$, then the line $t(z)$
cannot be a Cauchy line, as can be shown by some calculations summarized in the
following paragraph:

For a spacelike line $t(z)$ to be a Cauchy line in the flat Minkowski spacetime
(\ref{eq:41}), it must obviously have $z$ extend over the entire domain
$-\infty < z < \infty$, but this necessary condition is not sufficient.  One
must also have both of the standard null coordinates $u$ and $v$ extend over
their infinite range $-\infty < u < \infty$, $-\infty < v < \infty$, on the
line.  In terms of $p$ and $z$, one can readily calculate that along the line
$t = t(z)$,
 \begin{equation}
 u \equiv t - z = t(0) - \int_{0}^{z}{dz
 \left( 1 - {p \over \sqrt{1+p^2}} \right)},
 \label{eq:45}
 \end{equation}
 \begin{equation}
 v \equiv t + z = t(0) + \int_{0}^{z}{dz
 \left( 1 + {p \over \sqrt{1+p^2}} \right)}.
 \label{eq:46}
 \end{equation}
If $p$ oscillates indefinitely about 0 or remains bounded, both $u$ and $v$
will generally have an infinite range (though sufficiently asymmetric
oscillations which become unbounded can preclude this), and the line $t(z)$
will be a Cauchy line as $z$ covers the entire real axis.  However, if $p$
monotonically becomes arbitrarily large as $z$ tends to $\pm \infty$, one needs
the integral of $1/p^2$ to diverge.  If $p$ has a asymptotic power-law
dependence on $z$, then the exponent must be no larger than ${1 \over 2}$ in
order that the line be a Cauchy line.  On the other hand, the minimum
asymptotic exponent in order that the energy integrand in Eq. (\ref{eq:43})
remain nonpositive is ${2 \over 3}$, so the integrand cannot be positive over
an entire infinite range of $z$ if the line $t = t(z)$ is a Cauchy line.

In fact, one can show further that if $z_1$ is taken to $-\infty$ and $z_2$ is
taken to $+\infty$ along a Cauchy line, then the Killing energy between these
two limits cannot remain negative for the instantaneous vacuum corresponding to
this line.  This result might appear to follow simply from the nonnegativity of
the total energy in the unbounded Minkowski spacetime, but there is the logical
possibility that for a quantum state not in the same Fock sector as the
Minkowski vacuum, the energy in any finite region might be negative and only
compensated by a positive energy (probably necessarily infinite) that is
entirely at spatial infinity.  This possibility does indeed occur in 1+3
dimensional Minkowski spacetime, in which a regular state of a conformally
invariant scalar field with negative energy density everywhere has been found
by Brown, Ottewill, and Siklos \cite{BOS}.  However, the instantaneous vacuum
corresponding to a Cauchy line does not give a possible way of constructing
such a state, if one exists, for a massless field in 1+1 dimensional Minkowski
spacetime.

Neither does the 1+1 dimensional analogue of the 1+3 dimensional
negative-energy-density state \cite{BOS} give negative energy density
everywhere; instead, it gives precisely the ordinary vacuum state of zero
energy density everywhere in that case.  Some preliminary calculations suggest
to me (though I have do not have a rigorous proof) that in flat unbounded
Minkowski spacetime with 1+1 dimensions, unlike the case in 1+3 dimensions,
there are no regular states in which the first integral of Eq. (\ref{eq:40})
gives a negative total integrated `energy' $E$, even when one ignores a
possible contribution entirely at spatial infinity.

\section*{VII.  CONCLUSIONS}

\hspace{.25in}For a massless bosonic or fermionic field (possibly with
arbitrarily twisted boundary conditions) in 1+1 dimensional curved spacetime,
an extension of an elementary derivation of the trace anomaly given herein
shows that the instantaneous vacuum relative to any Cauchy line has a very
simple form for its stress-energy tensor on that line [Eqs.
(\ref{eq:7})--(\ref{eq:9}), or (\ref{eq:25})--(\ref{eq:27})].  This form
depends on the local properties of the line (its extrinsic curvature and its
derivative along the line) and of the spacetime there (its scalar curvature)
and on only one nonlocal quantity, a quadratic expression in the twisting of
the field, divided by the square of the length of the Cauchy line (assuming it
is closed; for an infinitely long Cauchy line this nonlocal quantity is zero).
The explicit form of the resulting stress-energy tensor can readily be used to
deduce various properties of instantaneous vacua in certain simple spacetimes,
such as the nonexistence of any quantum state with a finite deSitter-invariant
stress-energy tensor in covering spaces of the 1+1 dimensional deSitter
spacetime.

As asides to the basic calculations, the Casimir energy density for arbitrarily
twisted massless fields in static spacetimes that are periodic in one spatial
dimension is given [Eq. (\ref{eq:C})], and various covariant forms of the
generic stress-energy tensor of a massless quantum field in 1+1 dimension are
presented [Eqs. (\ref{eq:C1}), (\ref{eq:C9}), and (\ref{eq:C22})], in terms of
which one may find two nonlinear functionals [Eqs.
(\ref{eq:C16})--(\ref{eq:C17}) or (\ref{eq:C28})--(\ref{eq:C29})] that are
coordinate and conformal invariants of the quantum state.

\section*{ACKNOWLEDGMENTS}

\hspace{.25in}The research for this paper was motivated by my attempt to
understand a recent paper by Capri, Kobayashi, and Lamb \cite{CKL}.  (In
particular, it justifies their assumption that the energy density and flux are
zero for the instantaneous vacuum of a geodesic Cauchy line, but only when it
has infinite length.)  I am grateful to our newborn daughter Anna (born Dec.
12, 1995) for waking me up at 2 a.m., when, in order to be able do the
calculation in my head while trying to go back to sleep, I was led to come up
with the simple derivation of the trace anomaly given above.  Conversations
with Bruce Campbell, Valeri Frolov, Dmitri Fursaev, Chris Isham, Tom\'{a}\v{s}
Kopf, Pavel Krtou\v{s}, David Lamb, Bill Unruh, and Andrei Zelnikov helped me
clarify my comprehension and find references.  I am particularly grateful to
Werner Israel, whose suggestion to write the stress-energy tensor in covariant
form led to the elegant form for the coordinate and conformal invariants
determined nonlocally by the stress-energy tensor.  Financial support has been
provided by the Natural Sciences and Engineering Research Council of Canada.



\baselineskip 5pt

\begin{thebibliography}{99}

\bibitem{CD} D. M. Capper and M. J. Duff, Nuovo Cimento {\bf A23}, 173 (1974).

\bibitem{Duff} M. J. Duff, in {\em Quantum Gravity:  An Oxford Symposium}, eds.
C. J. Isham, R. Penrose, and D. W. Sciama (Clarendon Press, Oxford, 1975)

\bibitem{CPhD} S. M. Christensen, Ph.D. Dissertation, University of Texas
(unpublished, 1975).

\bibitem{DFU} P. C. W. Davies, S. A. Fulling, and W. G. Unruh, Phys.\ Rev.\
{\bf D13}, 2720 (1976).

\bibitem{FD} S. A. Fulling and P. C. W. Davies, Proc.\ R.\ Soc.\ Lond.\ {\bf
A348}, 393 (1976).

\bibitem{DDI} S. Deser, M. J. Duff, and C. J. Isham, Nucl.\ Phys.\ {\bf B111},
45 (1976).

\bibitem{DF} P. C. W. Davies and S. A. Fulling, Proc.\ R.\ Soc.\ Lond.\ {\bf
A354}, 59 (1977).

\bibitem{CF} S. M. Christensen and S. A. Fulling, Phys.\ Rev.\ {\bf D15}, 2088
(1977).

\bibitem{B} L. S. Brown, Phys.\ Rev.\ {\bf D15}, 1469 (1977).

\bibitem{F77} S. A. Fulling, J.\ Phys.\ A: Gen.\ Phys.\ {\bf 10}, 917 (1977).

\bibitem{W77} R. M. Wald, Commun.\ Math.\ Phys.\ {\bf 54}, 1 (1977).

\bibitem{D77} P. C. W. Davies, Proc.\ R.\ Soc.\ Lond.\ {\bf A354}, 529 (1977).

\bibitem{BD78} T. S. Bunch and P. C. W. Davies, Proc.\ R.\ Soc.\ Lond.\ {\bf
A360}, 117 (1978).

\bibitem{DU} P. C. W. Davies and W. G. Unruh, Proc.\ R.\ Soc.\ Lond.\ {\bf
A356}, 259 (1977).

\bibitem{W78} R. M. Wald, Ann.\ Phys.\ (N.Y.) {\bf 110}, 472 (1978).

\bibitem{BD} N. D. Birrell and P.C. W. Davies, {\em Quantum Fields in Curved
Space} (Cambridge University Press, Cambridge, 1982).

\bibitem{F} S. A. Fulling, {\em Aspects of Quantum Field Theory in Curved
Space-Time} (Cambridge University Press, Cambridge, 1989).

\bibitem{MTW} C. W. Misner, K. S. Thorne, and J. A. Wheeler, {\em Gravitation}
(W. H. Freeman, San Francisco, 1973).

\bibitem{C} H. B. G. Casimir, Proc.\ Kon.\ Ned.\ Akad.\ Wet.\ {\bf 51}, 793
(1948).

\bibitem{SS} B. Schroer and J. A. Swieca, Phys.\ Rev.\ {\bf D2}, 2938 (1970).

\bibitem{FPhD} S. A. Fulling, Ph.D. Dissertation, Princeton University
(unpublished, 1972).

\bibitem{I} C. J. Isham, Proc.\ R.\ Soc.\ Lond.\ {\bf A362}, 383 (1978).

\bibitem{I2} C. J. Isham, Proc.\ R.\ Soc.\ Lond.\ {\bf A364}, 591 (1978).

\bibitem{GSW} See, for example, M. B. Green, J. H. Schwarz, and E. Witten, {\em
Superstring Theory, Volume 1:  Introduction} (Cambridge University Press,
Cambridge, 1987).

\bibitem{R} P. Ramond, Phys.\ Rev.\ {\bf D3}, 2415 (1971).

\bibitem{NS} A. Neveu and J. H. Schwarz, Nucl.\ Phys.\ {\bf B31}, 86 (1971).

\bibitem{DHVW} L. Dixon, J. Harvey, C. Vafa, and E. Witten, Nucl.\ Phys.\ {\bf
B261}, 678 (1985); {\bf B274}, 285 (1986).

\bibitem{KLT} H. Kawai, D. C. Lewellen, and S.-H. H. Tye, Phys.\ Rev.\ Lett.\
{\bf 57}, 1832 (1986); {\bf 58}, 429 (1987); Nucl.\ Phys.\ {\bf B288}, 1
(1987).

\bibitem{LT} For a textbook treatment, see
D. L\"{u}st and S. Theisen, {\em Lectures on String Theory} (Springer-Verlag,
Berlin, 1989).

\bibitem{T} D. J. Toms, Phys.\ Rev.\ {\bf D21}, 928 (1980).

\bibitem{F53} R. P. Feynman, Phys.\ Rev.\ {\bf 91}, 1291 (1953).

\bibitem{M} T. Matsubara, Prog.\ Theoret.\ Phys.\ (Kyoto) {\bf 14}, 351 (1955).

\bibitem{FW} A. L. Fetter and J. D. Walecka, {\em Quantum Theory of
Many-Particle Systems} (McGraw-Hill, New York, 1971).

\bibitem{F72} R. P. Feynman, {\em Statistical Mechanics:  A Set of Lectures}
(Benjamin, New York, 1972).

\bibitem{Be} C. W. Bernard, Phys.\ Rev.\ {\bf D9}, 3312 (1974).

\bibitem{P} V. N. Popov, {\em Functional Integrals in Quantum Field Theory and
Statistical Physics} (D. Reidel, Dordrecht, 1983).

\bibitem{D} A. Das, {\em Field Theory:  A Path Integral Approach} (World
Scientific, Singapore, 1993).

\bibitem{DHI} B. S. DeWitt, C. F. Hart, and C. J. Isham, Physica {\bf 96A}, 197
(1979), and in {\em Themes in Contemporary Physics}, ed. S. Deser (North
Holland, Amsterdam, 1979).

\bibitem{DB} J. S. Dowker and R. Banach, J.\ Phys.\ A: Gen.\ Phys.\ {\bf 11},
2255 (1978).

\bibitem{AI} S. J. Avis and C. J. Isham, Nucl.\ Phys.\ {\bf B156}, 441 (1979).

\bibitem{AS} {\em Handbook of Mathematical Functions, with Formulas, Graphs,
and Mathematical Tables}, eds. M. Abramowitz and I. A. Stegun (Dover, New York,
1965), \S 23.1.18, p. 805.

\bibitem{GR} I. S. Gradshteyn and I. M. Ryzhik, {\em Table of Integrals,
Series, and Products:  Corrected and Enlarged Edition},
(Academic Press, San Diego, 1980), \S 9.622.

\bibitem{W} W. Israel, private communication.

\bibitem{BOS} M. R. Brown, A. C. Ottewill, and S. T. C. Siklos, Phys.\ Rev.\
{\bf D26}, 1881 (1982); {\bf D28}, 1560 (1983).

\bibitem{CKL} A. Z. Capri, M. Kobayashi, and D. J. Lamb, Class.\ Quant.\ Grav. {\bf 13}, 179 (1996), hep-th/9508077.

\end{thebibliography}
\end{document}